\documentclass[11pt,a4paper]{article}

\usepackage{jheppub}
\usepackage{axodraw}
\usepackage{graphics}

\author[a]{B C Allanach}
\affiliation[a]{DAMTP, CMS, University of Cambridge, Wilberforce Road, Cambridge CB3 0WA, United Kingdom}

\emailAdd{B.C.Allanach@damtp.cam.ac.uk}

\title{Impact of CMS Multi-jets and Missing Energy Search on CMSSM Fits}

\abstract{Recent CMS data significantly extend the exclusion limits for
  supersymmetry. We examine the impact of such data on global fits of the
  constrained minimal supersymmetric standard model (CMSSM) to indirect and
  cosmological data.
  By simulating supersymmetric signal events at the LHC, we construct
  a likelihood map for the recent CMS data, validating it against the
  exclusion region calculated by the experiment itself. 
  A previous CMSSM global fit is then re-weighted by our
  likelihood map. The CMS results nibble away at the high fit probability
  density region, 
  transforming probability distributions for the scalar and gluino masses.
The CMS search has a 
  non-trivial effect on 
  $\tan \beta$ due to correlations between the parameters implied by the fits
  to indirect data. 
}

\newcommand{\fourgraphs}[4]{%
\unitlength=1in
\begin{picture}(6,4)(0,0)
\put(-0.5,2){\epsfig{file={#1}.eps, width=3in}}
\put(-0.1,4){(a)}
\put(2.8,2){\epsfig{file={#2}, width=3in}}
\put(3.15,4){(b)}
\put(-0.5,0){\epsfig{file={#3}.eps, width=3in}}
\put(-0.1,2){(c)}
\put(2.8,0){\epsfig{file={#4}, width=3in}}
\put(3.15,2){(d)}
\put(1.5,1.45){\includegraphics[width=20pt]{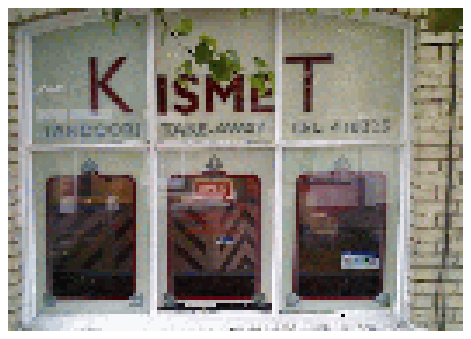}}
\put(1.5,3.45){\includegraphics[width=20pt]{kismet}}
\put(4.8,1.45){\includegraphics[width=20pt]{kismet}}
\put(4.8,3.45){\includegraphics[width=20pt]{kismet}}
\end{picture} 
}

\begin{document}
\maketitle

\section{Introduction}
Many recent efforts have examined whether various simple supersymmetric
models are compatible with the anomalous magnetic moment of the muon, the dark
matter relic density, direct searches for supersymmetric particles and Higgs
bosons, and electroweak observables simultaneously~\cite{Allanach:2005kz,Allanach:2006jc,Trotta:2006ew,Allanach:2006cc,Allanach:2007qk,Roszkowski:2007va,Allanach:2008iq,Feroz:2008wr,Buchmueller:2008qe,Trotta:2008bp,Roszkowski:2009sm,Feroz:2009dv,LopezFogliani:2009np,Roszkowski:2009ye,Buchmueller:2009fn,Buchmueller:2009ng,Baer:2010ny,Buchmueller:2010ai}. 
Global fits take into account variations with respect to all of the parameters
of the model that have a significant
impact on the observables, including Standard Model (SM) parameters. Various
algorithmic tools have now been developed 
to allow such a sampling of a multi-dimensional parameter space, which may be
multi-modal~\cite{Allanach:2007qj,Feroz:2008xx,Feroz:2011bj}. 
However, current global fits of indirect and cosmological data to the CMSSM
are not robust. This is shown by the large prior dependence of the Bayesian
fits~\cite{Allanach:2006jc,Allanach:2007qk,Trotta:2008bp,Allanach:2008iq}.
Frequentist fits to edge measurements from hypothetical LHC SUSY edge
measurements showed an
incorrect confidence level (C.L.)
coverage of frequentist fits when the C.L.s are calculated by assuming a
$\chi^2$ distribution~\cite{Bridges:2010de}. There are no published coverage
studies of global SUSY fits and, since they are expected to be less robust
than fits to an LHC SUSY signal, a coverage study of the frequentist fits is
necessary and long overdue. A fit to a large volume string model with only 
two free parameters additional to the SM (the ratio of the Higgs
vacuum expectation values, $\tan \beta$, and an overall supersymmetry breaking
mass scale) did display approximate prior independence~\cite{Allanach:2008tu}.  
On the other hand, a fit to a model with three parameters additional to the SM
(minimal anomaly
mediated supersymmetry breaking) showed significant prior
dependence~\cite{AbdusSalam:2009tr}.  
%
Fits to
models with more 
than three additional parameters have also (so
far) shown a lack of robustness~\cite{Roszkowski:2009sm,LopezFogliani:2009np,AbdusSalam:2009tr,AbdusSalam:2009qd}. 

The LHC general purpose experiments ATLAS and CMS actively searched for
supersymmetric states in 2010, 
collecting some $\sim 35$pb$^{-1}$ of recorded collisions each at a center of mass energy
$\sqrt{s}=7$ TeV. The first dedicated analysis of this data recently appeared
from CMS,
examining multi-jet states accompanied by missing transverse momentum~\cite{Khachatryan:2011tk}. 
The data were statistically compatible with the SM predictions,
allowing the experiment to place constraints on supersymmetry. 
In its publication, CMS also analyzed the CMSSM, showing that the equivalent
previous best 
exclusion from experiments at LEP and the Tevatron are significantly
surpassed. In 2011, some thirty times more luminosity is expected to be
collected by the LHC experiments than in 2010 and, in the absence of a signal,
the search reach will doubtless be extended
significantly. 

Here, we wish to examine the extent to which the first publicly available
search from CMS impacts on the
previous global supersymmetric fits. Despite the fact that such fits are not
yet robust and therefore not definitive (and are likely to remain so until direct
supersymmetric searches yield a significant signal), such an exercise should
give interesting information on whether the good-fit region is being covered
yet, and the prospects for searches in the near future. The effect of the CMS direct
SUSY search on the fits may be qualitatively 
examined. We demonstrate how one can perform such an analysis in a
reasonable amount of computational time, and hopefully set precedents for
good practice, such as validation against CMS's more sophisticated event
simulation.  
The CMS search has already been used to examine how much of the CMSSM
parameter space is allowed beyond the 95$\%$ exclusion
contour:~\cite{Strumia:2011dv}, around 1$\%$ according to the author's
metric related to the naturalness of electroweak symmetry breaking. 
We shall examine the effect of the search results on a previous Bayesian fit,
because of 
the computational ease with which the impact of the direct search can be 
quantified (as opposed to the frequentist methods, where an examination of
coverage would be necessary to set confidence levels, requiring many such
fits).  
We pick the CMSSM as our model partly because 
CMS analyzed it explicitly in their publication. Thus, it is possible to check 
our evaluation of the CMS data, which includes various approximations and
simplifications, against their more sophisticated treatment.
We use the 95$\%$ confidence
level exclusion contour in a two dimensional parameter plane that CMS provided
in their paper. 
The CMSSM is a
familiar model that many works have analyzed and it provides a well-defined
playground for examining supersymmetric phenomenology, with only a few free
parameters additional to the SM: the universal scalar mass $m_0$,
the universal gaugino mass $m_{1/2}$, the supersymmetric (SUSY) breaking
scalar trilinear 
coupling $A_0$, $\tan \beta$ and the sign of $\mu$, a parameter in the Higgs
potential. 
Although the CMSSM is very
specific and may well not represent the correct pattern of supersymmetry
breaking, aspects of its phenomenology are similar to the class of models which
are effectively perturbations around the CMSSM assumptions. 

Early ATLAS jets plus missing transverse momentum data presented at Summer
conferences in 2010 have already been used~\cite{Alves:2010za} to place bounds
on simplified 
models containing squarks and gluinos which decay to jets and neutralinos.
This analysis was based only on 70 nb$^{-1}$ of data. This data is not
used here because it is only a small integrated luminosity, with a small
effect compared to the data we use. 
We shall see that the predicted CMS SUSY search signal is approximately
independent of all parameters except $m_0$ and $m_{1/2}$, which leads to
significant simplifications in the incorporation of its results into the
fits. 

Our paper proceeds as follows: in section~\ref{sec:CMS}, we detail the CMS
SUSY search and results, showing how we approximately reproduce their
analysis, resulting in a likelihood map on the CMSSM parameter space. 
We then go on to apply the approximate likelihood map to global CMSSM fits in
section~\ref{sec:fits}, showing how the fits change.  We conclude in
section~\ref{sec:summ}. We calculate the accuracy of our approximation that
the CMS SUSY signal is approximately independent of all parameters except
$m_0$ and $m_{1/2}$ in Appendix~\ref{sec:app}.

\section{CMS $\alpha_T$ Likelihood Map \label{sec:CMS}}

\subsection{The $\alpha_T$ Search}
In 35pb$^{-1}$ of $pp$ collisions at $\sqrt{s}=7$ TeV at the CERN LHC, 
CMS examined events with significant transverse momentum  $|{\bf p_T}|$ in the
jets
 $j_i$ for multi-jet events, i.e.\ $H_T=\sum_{i=1}^{N_{jet}} |{\bf
  p}_T^{j_i}|>350$ GeV. ${\bf p}_T^{j_i}=(p_x^{j_i},\ p_y^{j_i})$ is the
jet momentum transverse to the beam. 
Isolated lepton and photon vetoes were also
applied. The final event selection relies on a variable $\alpha_T$, that is
designed to discriminate effectively against QCD multi-jet production, where
one of the jets' transverse momentum is significantly
mis-measured~\cite{Randall:2008rw}, although it has been argued that the
variable gives no special immunity to initial and final state radiation
effects~\cite{Usubov:2010gs}.  
The system of $N_{j}$ jets is reduced to a system of two jets by combining 
them into pseudo-jets $A$ and $B$\footnote{We 
  note here that the difference between transverse momentum and transverse
  energy can be significant for pseudo-jets, even within an event generator,
  because they have effective masses. Like the authors of Ref.~\protect\cite{Barr:2010zj}, we
  urge care when defining them.}.
The combination chosen is
the one 
that minimizes
\begin{equation}
\Delta H_T \equiv \sum_{j_i \in A} |{\bf p}_T^{j_i}| - \sum_{j_i \in B} |{\bf
  p}_T^{j_i}|.
\end{equation}
One then calculates 
\begin{equation}
\alpha_T=\frac{H_T - \Delta H_T}{2\sqrt{H_T^2 -  {H\!\!\!\!/}_T^2}}, \label{alpha}
\end{equation}
where ${H\!\!\!\!/}_T=\sqrt{(\sum_{i=1}^{N_{jet}} p_x^{j_i})^2+
(\sum_{i=1}^{N_{jet}} p_y^{j_i})^2}$ is defined at the jet-level. Note that often, a
different definition of $\alpha_T$ is given which is identical in the idealized
case of massless di-jet events, but which differs from the result of
Eq.~\ref{alpha} for the case of more than two jets.
The cut $\alpha_T>0.55$ is used in CMS's analysis.

The jets are defined by
the 
anti-$k_T$ algorithm with a size parameter of 0.5, a minimum $|{\bf p_T}|>50$ GeV and
a bound on the pseudo-rapidity $\eta$ of $|\eta|<3$. The highest $|{\bf p_T}|$
jet is 
required to have $|\eta|<2.5$ and the transverse momentum of the two leading
jets must exceed 100 GeV. 
${{\bf p}\!\!\!/}_T(cal)$ is defined over all calorimetric deposits of the
event, 
and events with $R_{\mbox{miss}}={H\!\!\!\!/}_T / |{{\bf p}\!\!\!/}_T(cal)| > 1.25$ are
rejected to protect against multiple jets failing the $|{\bf p}_T|>50$ GeV
requirement. 

Additional experimental cuts to remove events where significant energy losses
from jets going into un-instrumented regions were also placed. 
After all of these cuts, $o=13$ events on a predicted SM background of
9.3$\pm$0.9 were observed. This search is compatible with SM
predictions at the 2$\sigma$ level,
with a $\sim 1 \sigma$ slight excess of events.

\subsection{Simulation of the SUSY signal}
In order to simulate the production of sparticles from LHC $pp$ collisions, we use {\tt
  HERWIG++-2.4.2}~\cite{Bahr:2008pv}, with the default
underlying event model switched on. Supersymmetric spectra are generated with
{\tt SOFTSUSY3.1.7}~\cite{Allanach:2001kg}, with decay branching ratios and
widths calculated by {\tt   SDECAY}~\cite{Muhlleitner:2003vg}. The information 
about supersymmetric masses, mixings, couplings and decays is transferred
between the programs by the SUSY Les Houches
Accord~\cite{Skands:2003cj}. {\tt 
  fastjet-2.4.1}~\cite{Cacciari:2005hq} 
is used to define the jets in the anti-$k_T$ scheme with $R=0.5$ in the energy
recombination scheme. The cuts mimic those of the experiment and
are summarized in Eq.~\ref{cuts}:
\begin{equation}
H_T>350\mbox{~GeV}, \ |{\bf
  p}_T^{j_2}|>100\mbox{~GeV},\ \alpha_T>0.55,\ R_{\mbox{miss}}<1.25,\ 
l~\mbox{isolation}. \label{cuts}
\end{equation}
The lepton isolation veto (`$l$ isolation') is implemented as
follows\footnote{CMS uses a lepton isolation criterion involving tracks and
  energy deposits in calorimeter cells. Since we do not perform a detector
  simulation, we do not have access to simulated tracks and so
  we must deviate slightly from CMS's lepton isolation criteria, although our
  criteria are very similar in effect. 
CMS' lepton isolation criteria are: in $\Delta R<0.3$ around the lepton, the
lepton is considered isolated if 
the scalar $p_T$ sum of all of the tracks in the cone plus the $|{\bf p_T}|$s 
of the energy deposits in the calorimeter divided by the $|{\bf p}_T|$ of the
lepton candidate is larger than 0.15.}:
any events with leptons 
that have less than 10 GeV of $|{\bf p}_T|$ in a cone of $\Delta
R=\sqrt{(\Delta \eta)^2 + (\Delta \phi)^2}<0.3$ are vetoed, where $\phi$ is
azimuthal angle around the beam-line. 
For a given CMSSM parameter point predicting $s$ supersymmetric events passing
the cuts, 
we calculate the likelihood to be
\begin{equation}
{\mathcal L}= \frac{e^{-(s+b)} (s+b)^o}{o!} \label{likelihood}
\end{equation}
from the Poisson distribution for the central value of the predicted number of
signal $s$ plus
background events $b=9.3$. We neglect the small uncertainty of 0.9 on
$b$: in the next section, where we validate our calculation, we shall see
that the effect of our approximations (including this one) are reasonable.

Our simulation of the signal is cruder than the CMS analysis: it is only
leading order in QCD, whereas CMS's includes next-to-leading order factors for
the overall cross-section, and we have not
performed a dedicated detector simulation to convolute the signal kinematic
distributions with the detector response. Even had we included these effects,
it would be important to check our approximations by reproducing CMS's
calculated signal because, being outside the collaboration, we do not have
access to the full detector simulation. There is no need for us to validate
the SM background, since we may use CMS's predicted rates for it.

\subsection{Validation of the SUSY signal}
\begin{figure}\begin{center}
\unitlength=1.1in
\begin{picture}(3,2)(0,0)
\put(0,0){\includegraphics[width=3.3in]{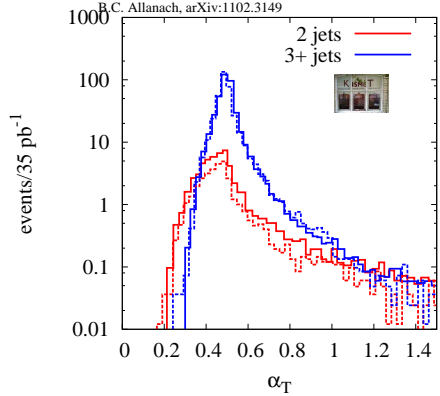}}
\put(2.02,1.45){\includegraphics[width=20pt]{kismet}}
\end{picture}
\caption{SUSY signal $\alpha_T$ distributions for the LM0 point and $\sqrt{s}=7$
  TeV in the di-jets channel and in the 3 or more jets channel, as displayed
  by the legend. Solid lines are obtained from Fig.~2 of
  Ref.~\protect\cite{Khachatryan:2011tk}: the CMS signal simulation 
  including next-to-leading order (NLO) corrections and full detector
  simulation whereas the dashed 
  lines show the results of our simulation and approximations.
The only cut applied is $H_T>350$ GeV.\label{fig:alphaTdist}}
\end{center}\end{figure}
In order to validate our calculation of the signal and see that we are
obtaining a reasonably accurate result despite our approximations, we
first calculate the $\alpha_T$ distribution in the di-jet and more than two jet
channels for the CMSSM model point LM0 ($m_0=200$ GeV, $m_{1/2}=160$ GeV,
$A_0=-400$ GeV, $\tan \beta=10$ and $\mu>0$). The only cut we place for the
purposes of this
check is $H_T>350$ GeV, mimicking CMS\@. The distributions for the `di-jets' and
`three or more jets' channels are shown in
Fig.~\ref{fig:alphaTdist} for our simulation and for CMS' signal simulation. 
The figure verifies that our calculation of the $\alpha_T$
distribution is
compatible with the calculation of the experimental collaboration: 
the
normalization of the sub-dominant exclusive two jets sample is slightly
different, but the shapes of both samples match extremely well. 
After all of the other cuts are applied, including $\alpha_T>0.55$, the
acceptance times efficiency of the SUSY signal selection is 5.0$\%$.

\begin{figure}\begin{center}{\unitlength=1in
\begin{picture}(6,2)(0,0)
\put(-0.5,0){\epsfig{file=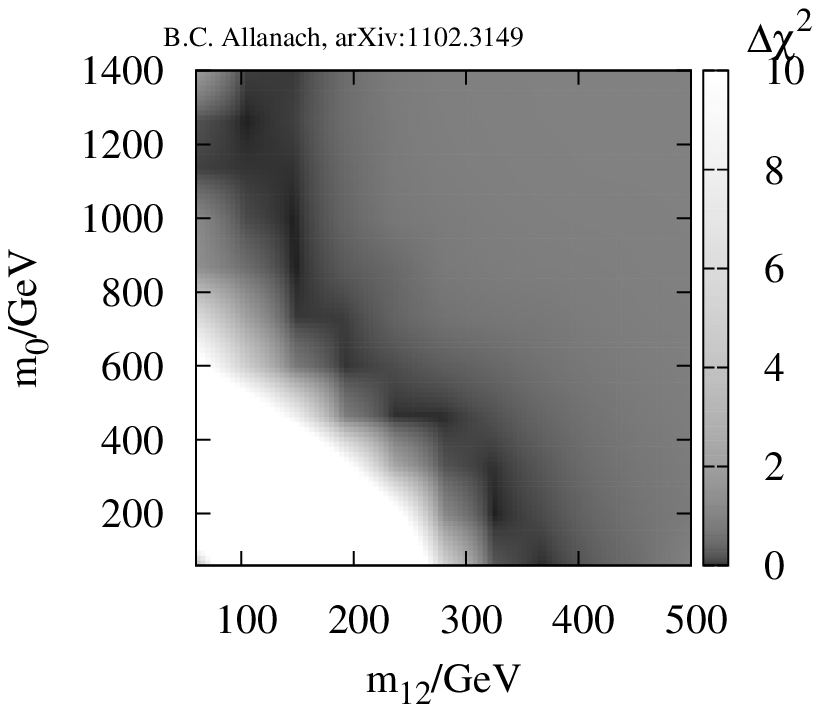, width=3.5in}}
\put(-0.05,2){(a)}
\put(0.125,0.44){\epsfig{file=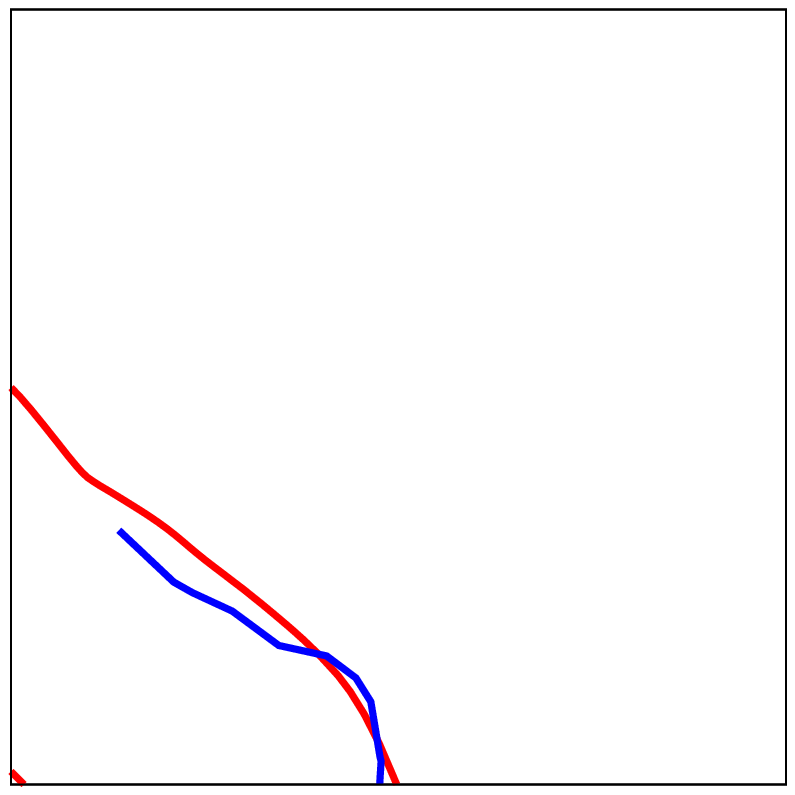, width=2.26in}}
\put(2.8,0){\epsfig{file=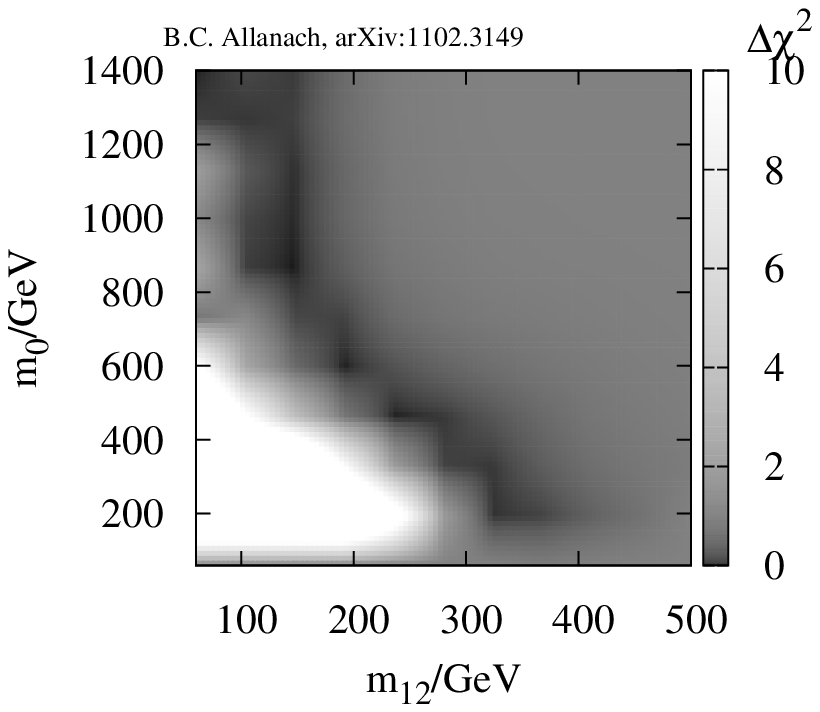, width=3.5in}}
\put(3.25,2){(b)}
\put(3.423,0.44){\epsfig{file=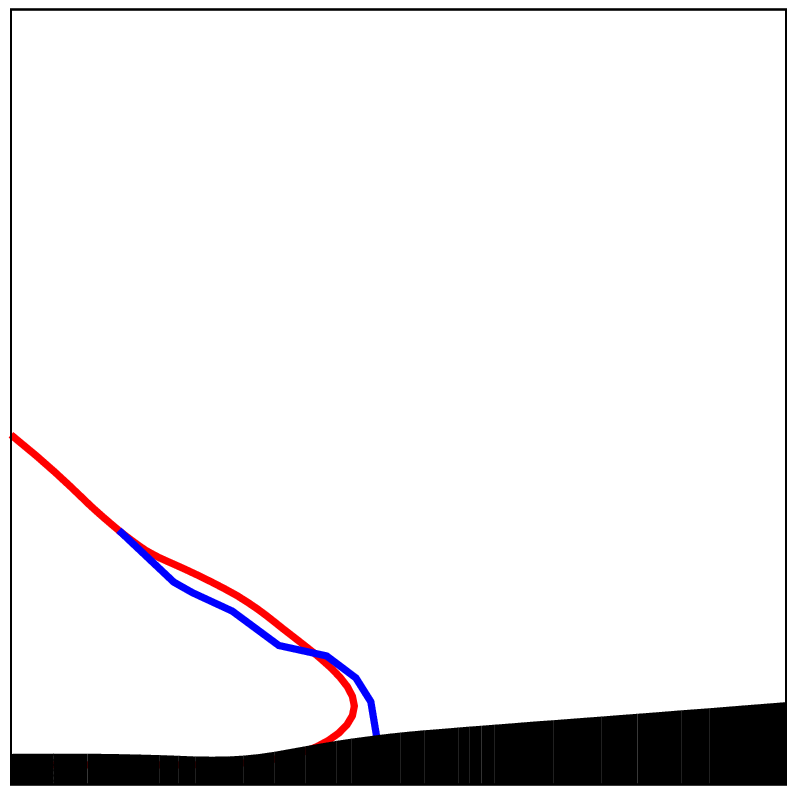, width=2.26in}}
\put(1.6,1.65){\includegraphics[width=20pt]{kismet}}
\put(4.9,1.65){\includegraphics[width=20pt]{kismet}}
\end{picture} 
\caption{Our approximation to the CMS $\alpha_T$-search CMSSM likelihood map
  for (a) 
  $\tan \beta=3$, $A_0=0$, 
  (b) 
  $\tan \beta=30$ and $A_0=-200$ GeV, where the blacked out  region at the
  bottom denotes a ${\tilde \tau}_1$ lightest supersymmetric particle. The
  region below the red (lighter) 
  curve is excluded at $95\%$ confidence level (C.L.), and $\Delta
  \chi^2$ is 
  clipped at 10. The CMS 95$\%$ C.L. curve is shown as the blue (darker)
  line. 
  \label{fig:scan}}  
}\end{center}\end{figure}
Next, we perform a scan over CMSSM parameter space to see how closely we can
reproduce CMS's calculation of the 95$\%$ contour. Like CMS, we choose $\tan
\beta=3$ and $A_0=0$ for this scan. At each point in an 11 by 11 grid, we
simulate 10 000 SUSY events, calculating ${\mathcal L}$ from
Eq.~\ref{likelihood}. We then calculate $\Delta \chi^2=-2 \ln {\mathcal
  L}/{\mathcal L}(\mbox{max})$, where ${\mathcal L}(\mbox{max})$ is the
maximum value of the likelihood over the plane.   
$\Delta \chi^2$ is then interpolated between the grid points, and shown in
Fig.~\ref{fig:scan}a. The background colour density displays $\Delta \chi^2$,
and is clipped at an upper value of 10. 
The light green solid curve is our predicted 95$\%$
C.L. lower bound, coming from the $\Delta \chi^2=5.99$ contour. 
We compare this with the blue (darker) line, the published CMS contour for the
full next-to-leading order observed limit. 
We see that
despite our approximations, the contours for 
the expected sensitivity and the 95$\%$ C.L.\ limits
agree reasonably well. Indeed,
next-to-leading order corrections to the production cross-section will tend to
increase the signal rate, whereas we expect detector effects to decrease it. 
Thus we expect that the two effects cancel to some extent. 
We conclude that we should be able to use the likelihood as calculated here,
although we note that we would really like more likelihood information from
the experiments far away from the 95$\%$ level contour so that we could validate
our simulation more thoroughly. The $\sim 1 \sigma$ excess of observed events
manifests as a dark valley of $\chi^2<1$ in Fig.~\ref{fig:scan}. This will
have an effect on the CMSSM global fit, as we shall see. 

For the $\alpha_T$ search, where the signal involves just jets and missing
transverse momentum, 
we expect the signal rate to be approximately independent
of $\tan \beta$ and $A_0$. This is because the signal is dominated by the
strong cross-sections of squark and gluino production, which do not depend
to any significant degree on those parameters. Third generation squark masses
do depend upon $A_0$ and $\tan \beta$ to some extent, but the SUSY production
cross-sections have a dominant component coming from the two lighter
generations of squark and the gluino. The decay cascades of the first two generations of
squarks and 
gluinos are likewise expected to be quite insensitive to $A_0$ and $\tan
\beta$, being dominated by strong processes until the decay into the
lightest neutralino. 
Being able to model the dependence of the CMS $\alpha_T$-search likelihood on
$m_0$ 
and $m_{1/2}$, 
while ignoring the effect of $A_0$ and $\tan \beta$ leads to
a significant simplification when we come to take it into account in our
global CMSSM fits.  
We check the assumption of independence of ${\mathcal L}$ with respect to
$A_0$ and $\tan \beta$ 
with another $m_0$-$m_{1/2}$ scan at a different
parameter point: $A_0=-200$ GeV and $\tan \beta=30$, for illustration. 
The result is shown in Fig.~\ref{fig:scan}b: the likelihood is similar to the
one in Fig.~\ref{fig:scan}a, and changing $A_0$ and $\tan \beta$ has not had a
significant effect. Other constraints on parameters, such as those coming from 
having a charged lightest supersymmetric particle (shown by the blacked out
region at the bottom of Fig.~\ref{fig:scan}b) do display a significant $\tan
\beta$ and $A_0$ dependence, but these are already taken into account within
our fits and will not pose a problem. The very bottom of the plot has negative
mass squared staus, and the signal rate could not be reliably calculated
there. This has a slight effect on the 
interpolated 95$\%$ C.L.\ contour, and is responsible for it having an
apparent (false) turning 
point in 
$m_{1/2}$.  We shall quantify the effects of the $\tan \beta-A_0$ independence
assumption in our fits in Appendix~\ref{sec:app}.

For increasing $m_{1/2}$, we see $\Delta \chi^2$ reaching a 
constant in Fig.~\ref{fig:scan} because there is no SUSY signal, since squarks
and gluinos become too
heavy to be produced. At large $m_0$ and small $m_{1/2}$, the SUSY signal is strongly dominated by
gluino pair 
production, where the gluinos have three-body decays into squarks. Thus the
dependence of ${\mathcal L}$ on $m_0$, if it is above 1400 GeV, is negligible. 
We shall therefore model the likelihood as follows: we use $s=0$ for
$m_{1/2}>500$ GeV. For $m_0>1400$ GeV and $m_{1/2}<500$ GeV, we use the
${\mathcal L}$ value given by the $m_0=1400$ GeV line on the figure. For
$m_0<1400$ GeV and $m_{1/2}<500$ GeV, we interpolate linearly within the grid of
likelihoods calculated. 

\section{Global CMSSM Fits Including the CMS $\alpha_T$ Search \label{sec:fits}}
We shall use a previous global Bayesian fit of the CMSSM from the {\tt KISMET}
collaboration~\cite{Allanach:2007qk}
to: the relic density of dark matter, the anomalous magnetic moment of the
muon, previous direct searches for sparticles, the branching ratios $BR(b
\rightarrow s \gamma)$, $BR(B_s \rightarrow \mu\mu)$, $M_W$, $\sin^2
\theta_w^l$, as well as 95$\%$ exclusions from LEP and Tevatron direct search
data. 
In order to predict $BR(b \rightarrow s \gamma)$, $BR(B_s \rightarrow
\mu\mu)$, the dark matter relic density and the anomalous magnetic 
moment of the muon, {\tt
  micrOMEGAs1.3.6}~\cite{Belanger:2001fz,Belanger:2004yn} was employed. It
computes $BR(b 
\rightarrow s \gamma)$ including one-loop SUSY corrections and NLO QCD
corrections. $BR(B_s \rightarrow \mu\mu)$ is calculated to one-loop. 
{\tt micrOMEGAs} calculates the non Standard Model correction to the
anomalous magnetic moment of the muon to one loop, and
we added the dominant two-loop SUSY QED
contributions from Ref.~\cite{Heinemeyer:2003dq,Heinemeyer:2004yq}. {\tt
  SUSY-POPE}~\cite{Heinemeyer:2006px}, which includes SUSY contributions up to
two loops, was used to calculate  
$M_W$ and $\sin^2 
\theta_w$.
Spectral predictions from {\tt SOFTSUSY3.1.7} are used to place bounds from
pre-LHC direct searches for sparticles and the lightest CP-even higgs boson.
{\tt SOFTSUSY3.1.7} calculates the sparticle masses to two-loop in the
renormalisation group equations, and one-loop in the threshold corrections. 
The dominant two-loop MSSM corrections are added to the Higgs mass
computation, and a 2 GeV error on the prediction of $m_h$ was assigned. 

The ranges of input parameters considered were: $2 < \tan \beta < 62$,
$|A_0|/\mbox{TeV}<4$, $60<m_{1/2}/\mbox{GeV}<2000$, $60<m_0/\mbox{GeV}<4000$. 
Variations with respect to the top mass $m_t$, the strong coupling constant $\alpha_s(M_Z)$,
the fine structure constant $\alpha(M_Z)$ and the bottom quark mass $m_b(m_b)$
were all included. Various 
different prior distributions were examined in Ref.~\cite{Allanach:2007qk},
but here we 
shall use the example of priors flat in the parameters listed above, except
for $m_0$ and $m_{1/2}$, which are flat in their logarithm. Using such log
priors allows us to illustrate the effects of the $\alpha_T$ search more
acutely than with purely linear priors. Rigorous convergence criteria were
satisfied by the fits, which were performed by ten Metropolis Markov Chains
running simultaneously. For more details on the fits, we refer interested
readers to Ref.~\cite{Allanach:2007qk}.

The effect of the data
is encoded within a likelihood function $L(m)$ defined on model parameters
$m$, and Bayesian inference consists 
of turning a prior probability distribution $\pi(m)$ into a posterior
probability distribution $p(m|d)$ via Bayes' theorem $p(m|d) = L(m) \pi(m)/D$.
$D=\int dm\ L(m) \pi(m)$ is a constant over parameter space: the Bayesian
evidence.  Since we
are performing parameter estimation, we are never interested in normalizing
constants, because the inference is performed under the hypothesis that the
model we are examining is correct and therefore the constants are always
normalized such that the total posterior probability is 1.
We take 2.7 million points from the Markov chain fits. The density of sampled
points is proportional to the posterior probability density at any point.
Each point $i$ typically appears sampled several times, and we therefore
set its weight 
$\rho_G(p_i)$ equal to the number of times it appears.
We write the point's parameters as $$p_i = \{ m_0,\ A_0,\ m_{1/2},\ \tan
\beta, m_t, m_b(m_b), 
\alpha(M_Z), \alpha_s(M_Z)\}_i.$$ 
The Markov chain Monte Carlo statistically gives us proportionality of the
weight with the
posterior probability distribution $p_G(p_i)$ ({\em
  prior} to the inclusion of LHC data) of the point: $p_G(p_i) = k \rho_G(p_i)$,
where $k$ is some constant which does not depend on $p_i$.

In order to take into account any new independent
data with likelihood $L_{new}(p_i)$, we have the combined likelihood
$L_{c}(p_i)=L_{new}(p_i) L_G(p_i)$ which combines the global fit data with the
new data.   Thus the combined posterior probability density function is 
$p_c(p_i) = c p_G(p_i) L_{new}(p_i)$, where $c$ is a constant that takes into
account the fact that the Bayesian evidence changes through the introduction
of new data. 
In terms of the Markov chain Monte Carlo point list, we obtain that 
$p_c(p_i) = ck \rho_G(p_i) L_{new}(p_i)$.
Ignoring the constants, we therefore calculate the new posterior by
re-weighting each point, multiplying its 
weight by the likelihood of the CMS $\alpha_T$ search in
Eq.~\ref{likelihood}. 
By plotting the posterior probability distributions before and after the
$\alpha_T$-re-weighting, we can 
then examine the effect of the $\alpha_T$ exclusion data on the CMSSM
fits. 

\begin{figure}\begin{center}{\unitlength=1in
\begin{picture}(6,2)(0,0)
\put(-0.5,0){\epsfig{file=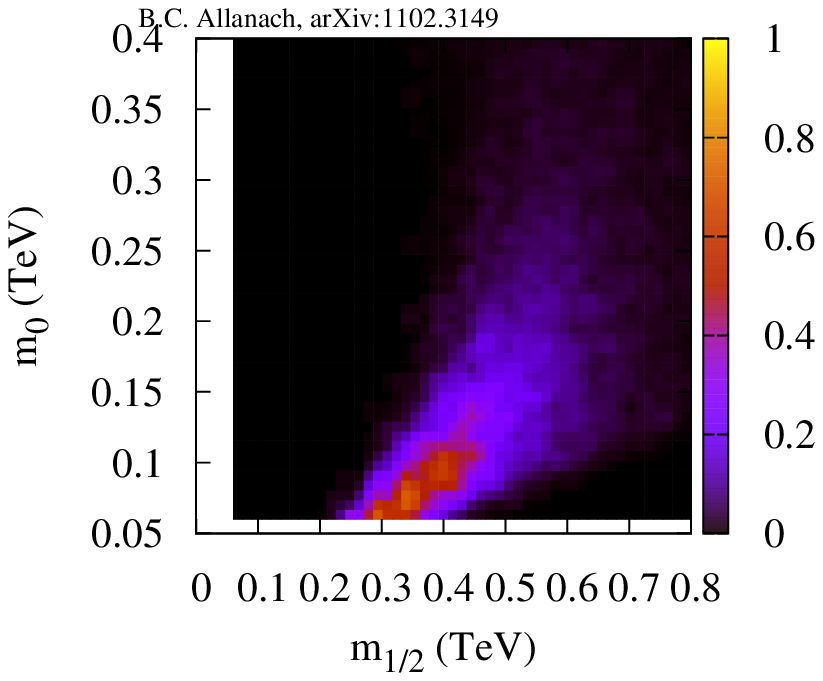, width=3.5in}}
\put(-0.05,2){(a)}
\put(2.8,0){\epsfig{file=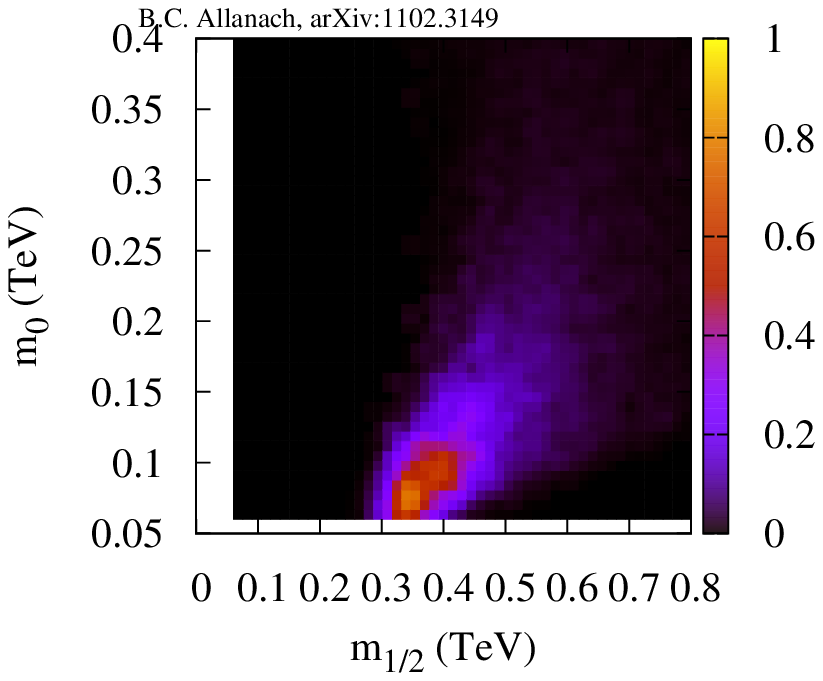, width=3.5in}}
\put(3.25,2){(b)}
\put(1.67,0.54){\includegraphics[width=20pt]{kismet}}
\put(4.97,0.54){\includegraphics[width=20pt]{kismet}}
\put(3.423,0.44){\epsfig{file=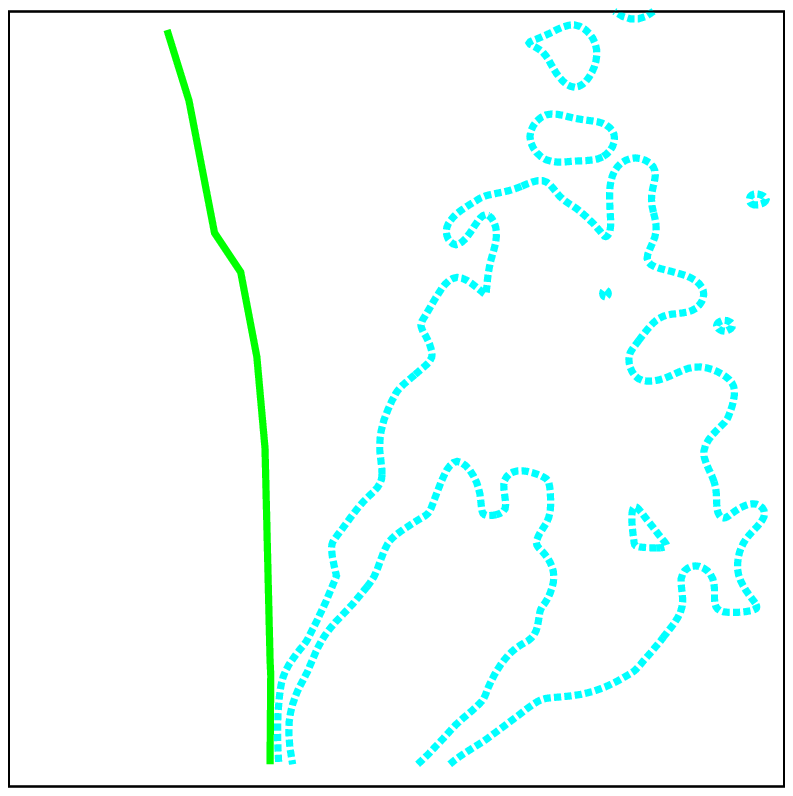, width=2.26in}}
\put(0.125,0.44){\epsfig{file=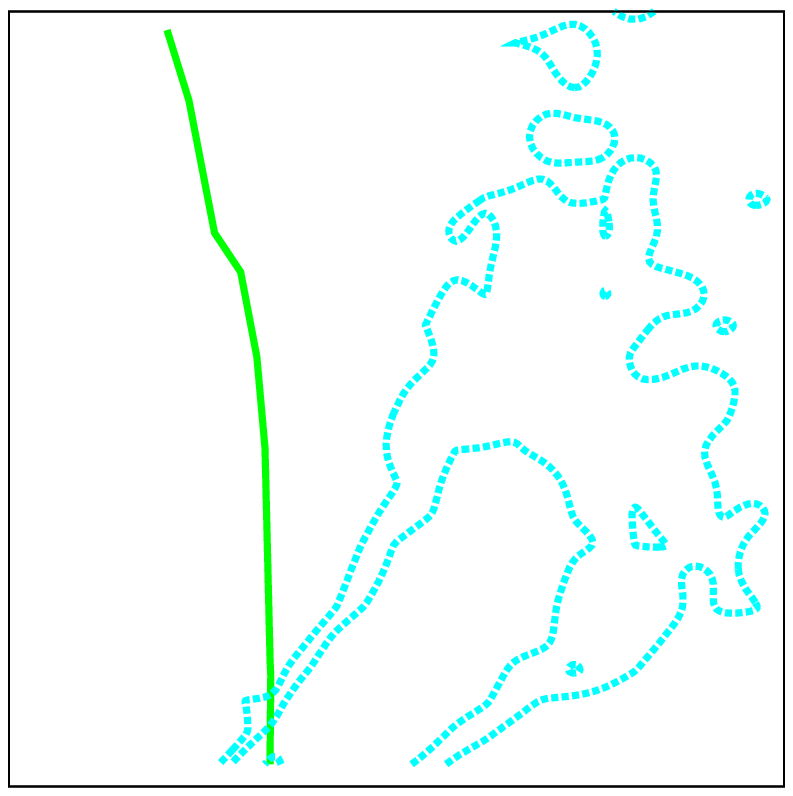, width=2.26in}}
\end{picture} 
\caption{Global CMSSM fits: (a) excluding the CMS $\alpha_T$ search and 
  (b) including the CMS $\alpha_T$ search likelihood. 
The posterior probability of each bin is shown as the background colour,
normalized to the maximum bin probability.
The almost vertical
curve shows our approximation to the CMS 95$\%$ exclusion. The dotted cyan inner
(outer) contour 
shows the 68$\%$ (95 $\%$) Bayesian credibility region. \label{fig:m0m12}}  
}\end{center}\end{figure}
We show the fits marginalized over all parameters except for $m_0$ and
$m_{1/2}$ in Fig.~\ref{fig:m0m12}. 
In order to guide the eye, we have added the 95$\%$ exclusion contour (the
light, almost vertical curve), but we remind the reader that the $\alpha_T$
likelihood was taken into account, not just a veto from this curve. 
Nevertheless, we see in the figure the expected behavior: there is a small
volume of good fit towards the bottom left of the plot that is ``chopped off''
by the $\alpha_T$ exclusion limit. There are also non-trivial effects on the
parameter plane: we see from Fig.~\ref{fig:scan}a that there is a slight $\sim
1\sigma$ preference of
the data against high $m_{1/2}$, since there was an excess of observed events
versus expected background. Such effects are present in the plot, but are easier
to see when the posterior probability density is marginalized down to one
dimension only (see below).

\begin{figure}\begin{center}{\unitlength=1in
\begin{picture}(6,2)(0,0)
\put(-0.5,0){\epsfig{file=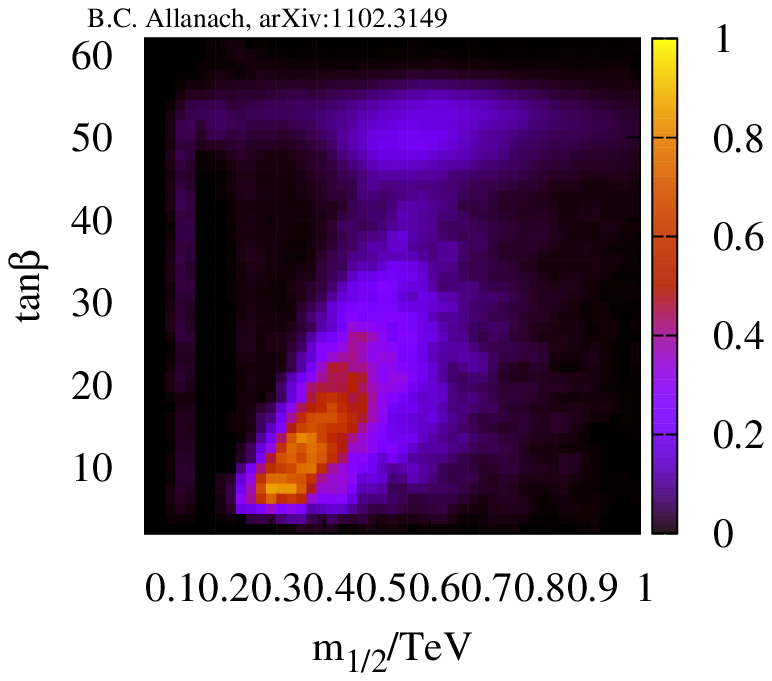, width=3.5in}}
\put(-0.05,2){(a)}
\put(2.8,0){\epsfig{file=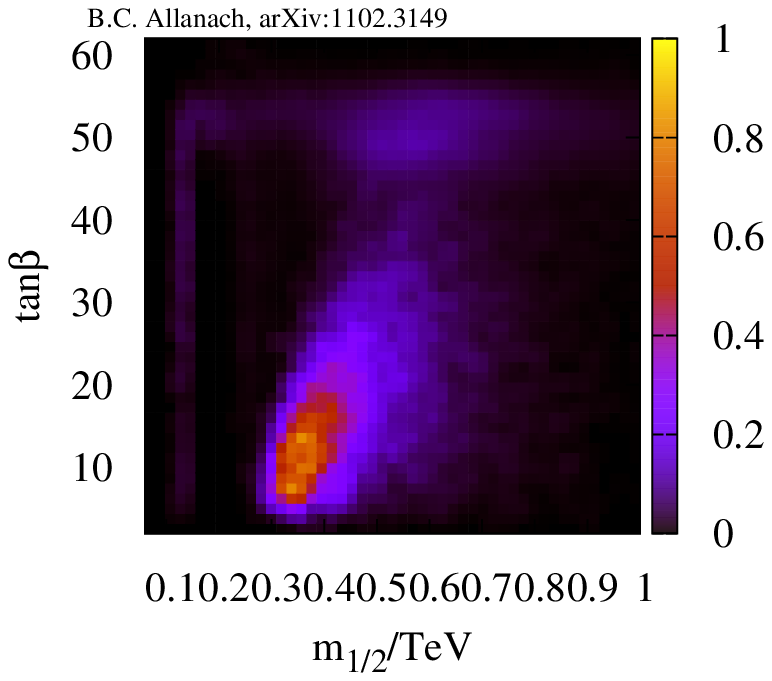, width=3.5in}}
\put(3.25,2){(b)}
\put(1.67,0.54){\includegraphics[width=20pt]{kismet}}
\put(4.97,0.54){\includegraphics[width=20pt]{kismet}}
\put(0.125,0.44){\epsfig{file=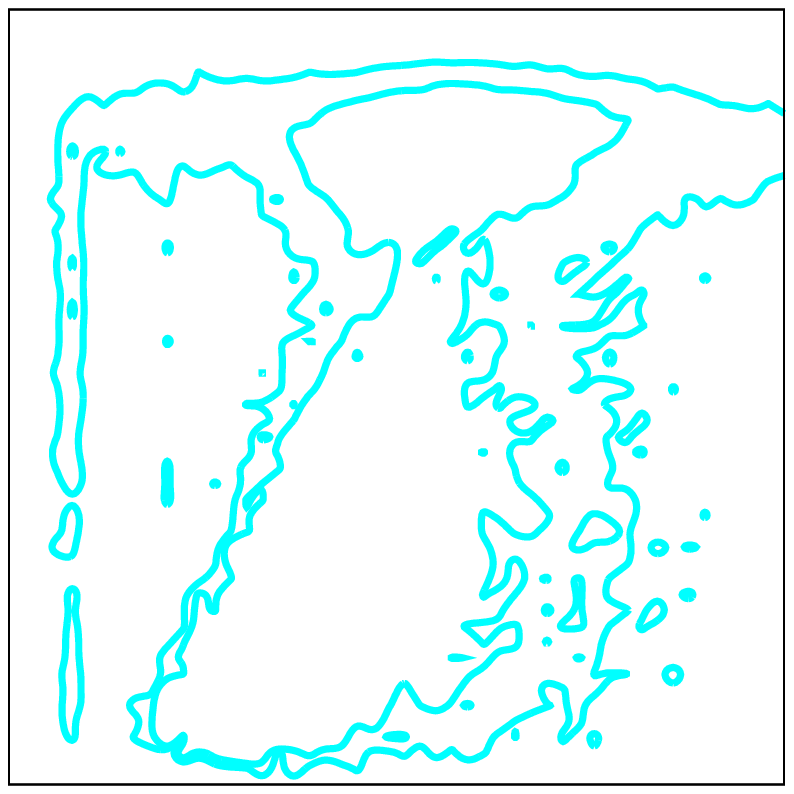, width=2.26in}}
\put(3.423,0.44){\epsfig{file=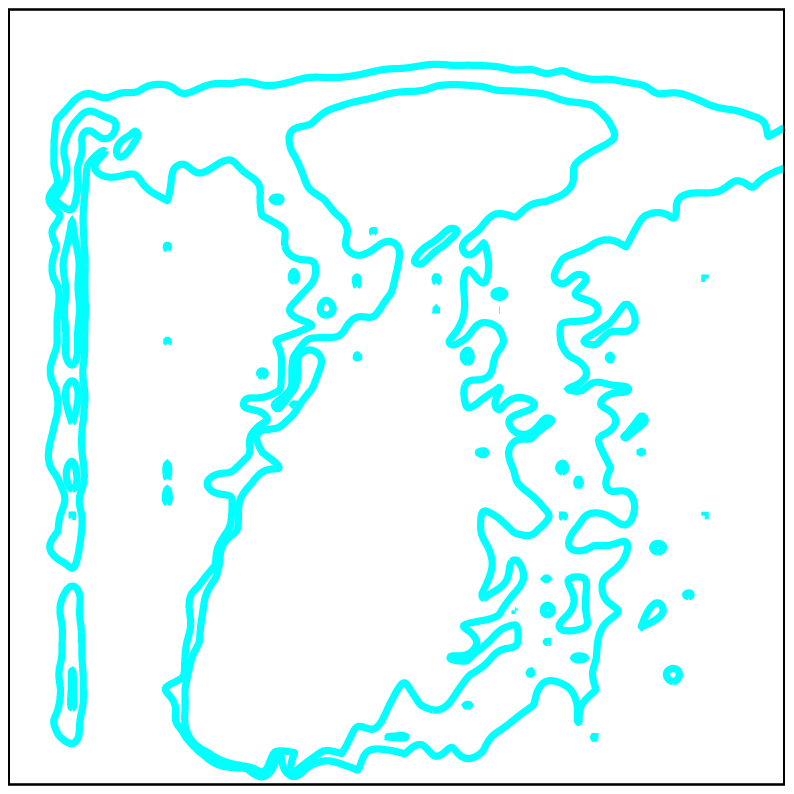, width=2.26in}}
\end{picture} 
\caption{Global CMSSM fits: (a) excluding the CMS $\alpha_T$ search and 
  (b) including the CMS $\alpha_T$ search likelihood. 
The posterior probability of each bin is shown as the background colour,
normalized to the maximum bin probability.
The dotted cyan inner
  (outer) contour 
shows the 68$\%$ (95 $\%$) Bayesian credibility region. \label{fig:m12tb}}  
}\end{center}\end{figure}
We show the marginalization to the $m_{1/2}-\tan \beta$ plane in
Fig.~\ref{fig:m12tb}. Here, we start to see the effect of non-trivial
correlations of the CMSSM parameters from the indirect fits. The long, thin
vertical region on the left-hand side of the figure corresponds to the
$h-$pole region, where $m_{\chi_1^0} \approx m_h/2$ and the lightest
neutralinos  ${\chi_1^0}$
efficiently annihilate in the early universe through the $s-$channel higgs ($h$)
pole~\cite{Drees:1992am}. This region is at large values of $m_0>1400$ GeV and
low 
$m_{1/2}$~\cite{Djouadi:2005dz}.  
Examining Fig.~\ref{fig:scan}a, we see that this region is given a likelihood
enhancement by the $\alpha_T$ search, and we see evidence for this in
Fig.~\ref{fig:m12tb}b: it gains some 68$\%$ level contours as its probability
mass increases. Also, the region at high $\tan \beta$ within the 68$\%$ level
contour has some of its probability mass removed, since it is at large values
of $m_{1/2}$, where there is a slight $\chi^2$ penalty compared to smaller
values of $m_{1/2}<400$ GeV, due to the $\alpha_T$ 1$\sigma$ excess of events. 

\begin{figure}
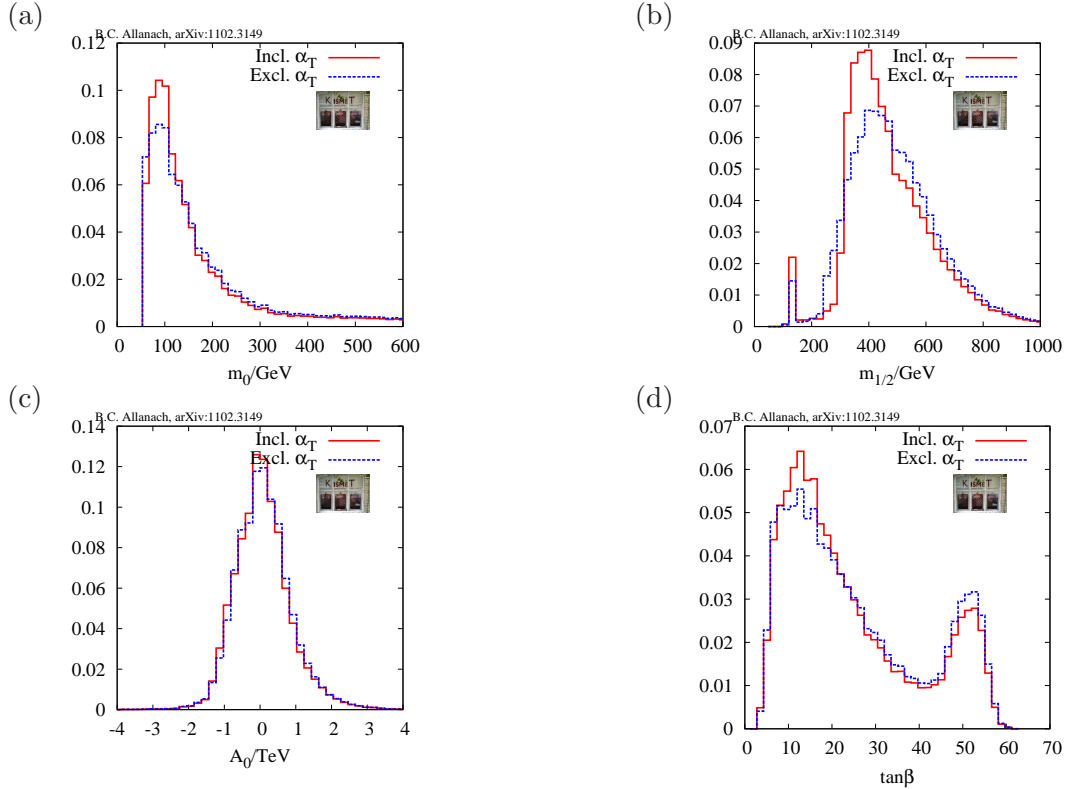
\begin{center}{\fourgraphs{m0}{m12}{A0}{tb}
\caption{Effect of the $\alpha_T$-search on one dimensional 
  probability distributions of CMSSM parameters. The area of each histogram has been
  normalized to 1 and labeled `Incl.\ $\alpha_T$' (`Excl.\ $\alpha_T$') if it
  includes (excludes) the CMS $\alpha_T$ search.
 \label{fig:params}}}\end{center}\end{figure}
We display the probability distributions of the individual CMSSM parameters
marginalized over all other parameters in Fig.~\ref{fig:params}. We see from
Fig.~\ref{fig:params}a that, because of correlations between the parameters in
the global fit, the
$\alpha_T$-search slightly prefers values of $m_0 \sim 100$ GeV. 
This is because the small $m_{1/2}$ region is boosted by the $1\sigma$ excess
in the number of observed events, which also has a small $m_0$ in order to
obtain sufficient neutralino annihilation in the early universe (in the stau
Co-annihilation region). 
Fig.~\ref{fig:params}b
shows that small values of $m_{1/2}$ are relatively disfavored by the
$\alpha_T$ search, 
except for the spike at $m_{1/2} \approx 120$ GeV which corresponds to the
$h-$pole region and is enhanced by including the $\alpha_T$ search. Intermediate values of $m_{1/2}$ of $300-500$
are also relatively favored compared to larger values. The $A_0$ distribution
is relatively untouched by the search, as shown in Fig.~\ref{fig:params}c, whereas
intermediate values of $\tan \beta=8-20$ are slightly preferred by it: 
see Fig.~\ref{fig:params}d.

\begin{figure}
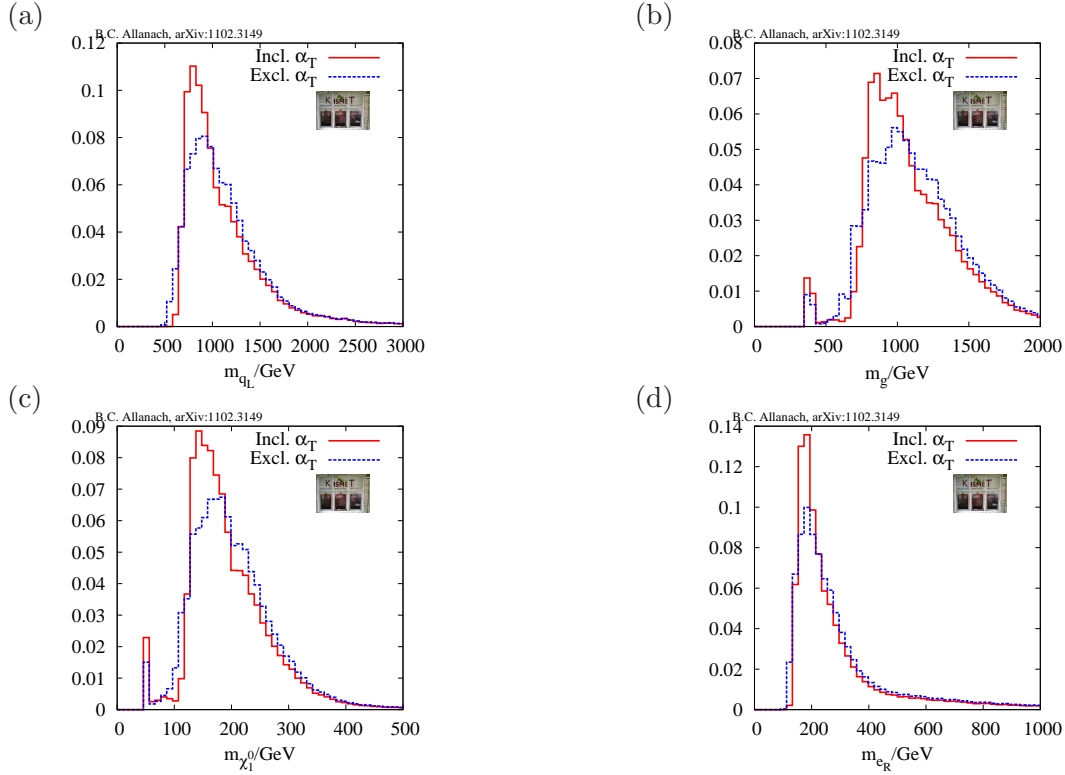
\begin{center}{\fourgraphs{msq}{mgl}{mchi10}{mslep}
\caption{Effect of the $\alpha_T$-search on the 
  probability distributions of sparticle masses in the CMSSM\@. The area of each
  histogram has been 
  normalized to 1 and labeled `Incl.\ $\alpha_T$' (`Excl.\ $\alpha_T$') if it
  includes (excludes) the CMS $\alpha_T$ search. \label{fig:masses}}  
}\end{center}\end{figure}
Next, we turn to the effect of the $\alpha_T$ search on sparticle masses. We
plot the posterior probability distributions of a representative sample of
four different masses in Fig.~\ref{fig:masses}. The right handed squark mass
$m_{q_R}$ 
is squeezed somewhat toward intermediate values of 600-1200 GeV 
by the 
$\alpha_T$ search, as shown in Fig.~\ref{fig:masses}a. 
A similar effect is found in the right-handed selectron mass $m_{e_R}$, which
is squeezed somewhat toward intermediate values of 160-200 GeV, as shown in
Fig.~\ref{fig:masses}d.  
$m_{\chi_1^0}$ and the gluino
mass $m_{g}$ are correlated strongly with $m_{1/2}$ and
Figs.~\ref{fig:masses}b,c show that the overall features follow those of the
$m_{1/2}$ distribution: a spike corresponding to the
$h-$pole at light masses and an enhancement of 
intermediate values of the masses to the detriment of more 
extreme values. 

\begin{figure}\begin{center}
\unitlength=1.1in
\begin{picture}(3,2)(0,0)
\put(0,0){\includegraphics[width=3.3in]{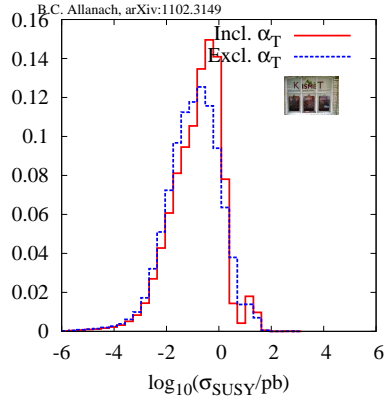}}
\put(2.02,1.45){\includegraphics[width=20pt]{kismet}}
\end{picture}
\caption{Effect of the $\alpha_T$-search on the total SUSY cross-section
  $\sigma_{SUSY}$ in the CMSSM in $pp$ collisions at $\sqrt{s}=7$ TeV. The
  area of each 
  histogram has been 
  normalized to 1 and labeled `Incl.\ $\alpha_T$' (`Excl.\ $\alpha_T$') if it
  includes (excludes) the CMS $\alpha_T$ search. 
  \label{fig:sigma}}
\end{center}\end{figure}
The probability distributions of masses found
predict a probability distribution for the cross-section of sparticle
production at the LHC\@. 
Approximately 1 fb$^{-1}$ of integrated luminosity is expected to be collected
over the next year at the LHC\@.  Thus, a ``weather forecast'' of whether it
will be raining 
sparticles (or whether we shall find a SUSY desert), is of special interest. We calculate
the total sparticle production cross-section $\sigma_{SUSY}$ predicted by our
fits at the LHC, with no cuts or modelling of detector effects. This quantity
then provides a crude upper bound on the actual SUSY signal cross-section that
may be observed. However, we do not attempt here to quantify the probability
of SUSY discovery in the next year, since such an inference is not yet robust due
to the insufficient constraining power of the data.  It appears from
Fig.~\ref{fig:sigma} that the prospects for SUSY discovery next year
marginally improve after the inclusion of the $\alpha_T$ search results.

\section{Summary and Conclusions \label{sec:summ}}
The CMS $\alpha_T$ search has significantly extended the previous exclusion
limits in the CMSSM\@. We have examined its effect on global fits of the CMSSM
to 
the anomalous magnetic moment of the muon, the dark
matter relic density  and electroweak observables while taking into account
previous direct searches for supersymmetric particles and Higgs
bosons.
The search 
nibbles 
away at the part of the parameter space where both squarks and gluinos are
light, but also has 
some other non-trivial effects. Because the search saw a slight excess,
intermediate masses acquire a small relative preference, with associated
effects on $\tan \beta$ because of parameter correlations in the global fit.

It is remarkable that even with the crudest of detector simulations
(consisting of $p_T$ and $\eta$ cuts), and leading-order event generation,
our 95$\%$ exclusion contour in a certain CMSSM parameter plane is close
to CMS's more sophisticated treatment. However, in order to be sure that we
approximately reproduce the CMS likelihood elsewhere in parameter space, we
require more details from the experimental publication. 
{\em We strongly advocate that the experiments publish more detail in order
  to allow more detailed checks: at 
 least additional confidence level contours or, even better, the
likelihood across a 
parameter plane, or ideally a {\tt RooStats} workspace}.

As this work was being completed,
ATLAS produced a search for jets, one lepton and missing
transverse momentum~\cite{Collaboration:2011hh} which extends the exclusion in
the CMSSM plane as compared 
to the CMS search that we consider. The ATLAS search has an observed number of
events 
lower than the expectation due to SM background (although the
measurement is statistically compatible with the background). 
We leave the inclusion of this data
to a future publication: it is likely to be  computationally
more difficult to include in our
fits than the ones contained in the present paper because the
signal rate may well have a dependence on all of the CMSSM parameters.
Event simulation is a CPU-time bottle neck, and if it depends on more than
just two or three parameters, is likely to need to be repeated for
different sampled fit points since it couldn't be easily interpolated in
a reasonable amount of CPU time anyway. 
The ATLAS data will thus require other more computationally intensive
techniques to analyze. 
Some of the effects that we have observed from the $\alpha_T$ CMS search, such
as an enhancement of intermediate values of sparticle masses due to the slight
excess of 
events, will presumably be reversed when the recent ATLAS data is taken into
account. 

\appendix

\section{Validation of $\tan \beta-A_0$ Independence of the Likelihood\label{sec:app}}

\begin{figure}\begin{center}
\unitlength=1in
\begin{picture}(6,2)
\put(0,0){\epsfig{file=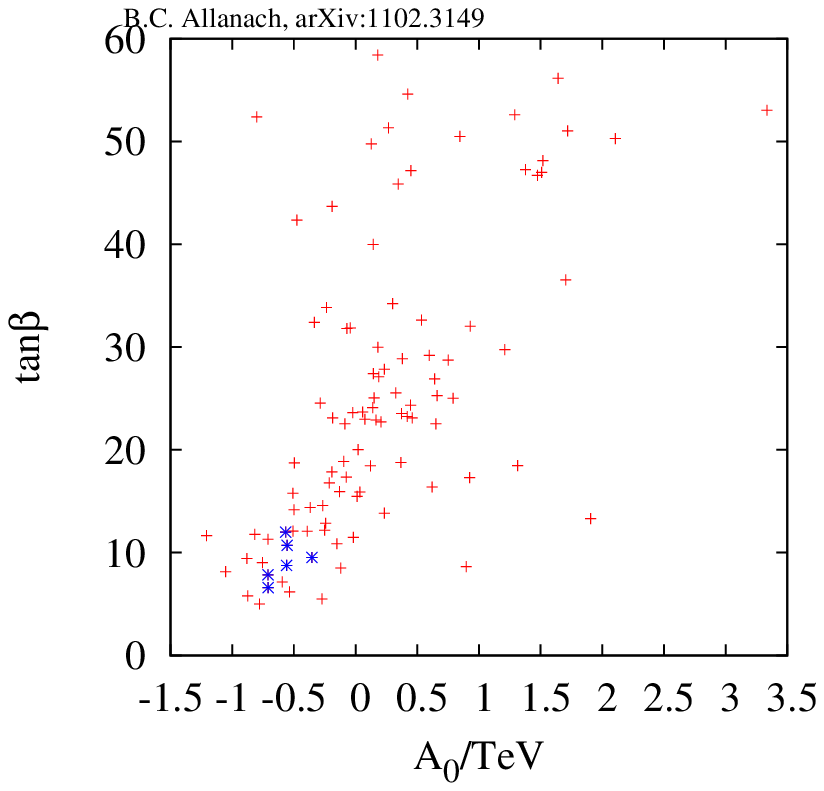,height=2.3in}}
\put(3,0){\epsfig{file=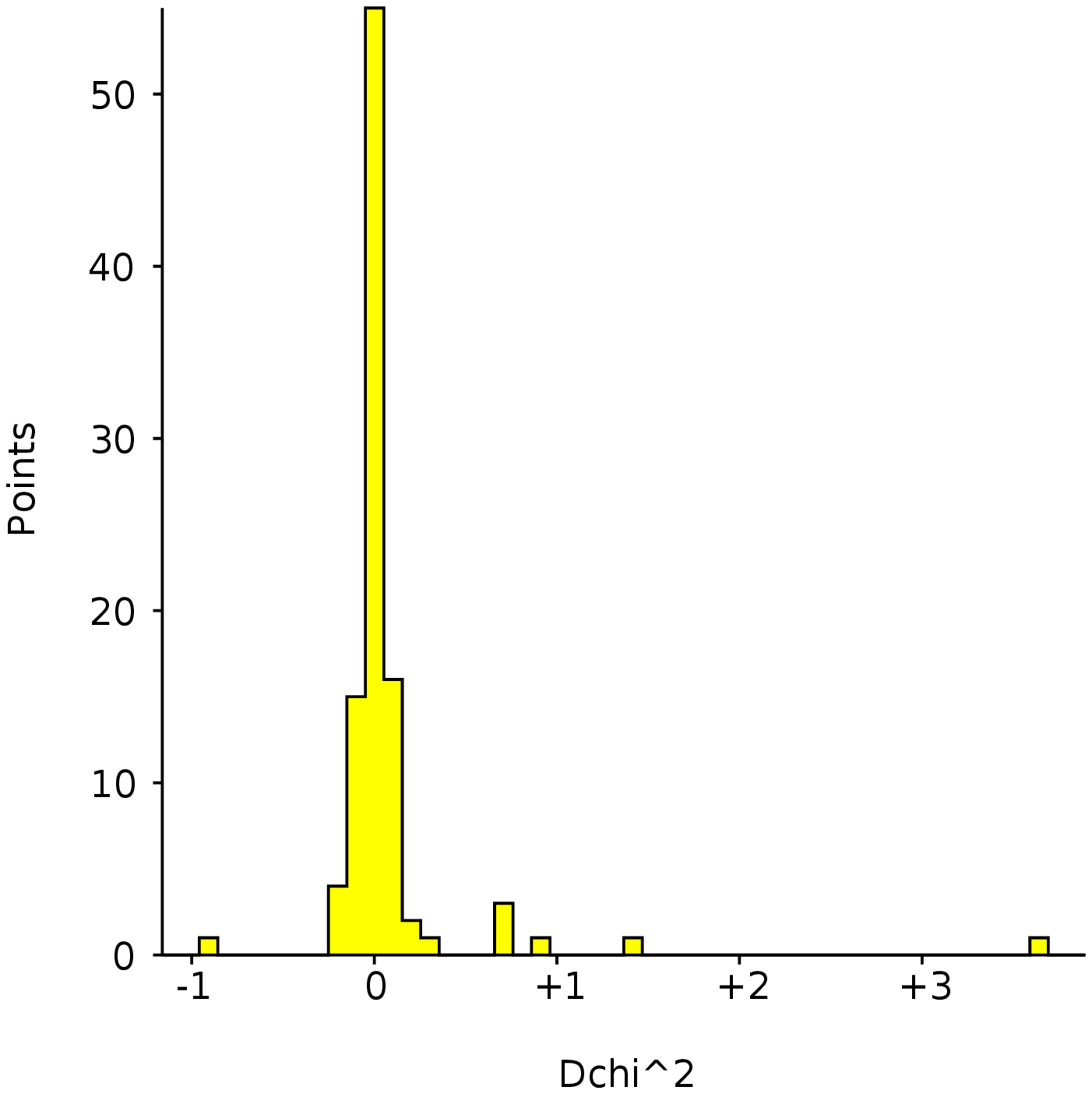,height=2.3in}}
\put(0,2){(a)}
\put(3,2){(b)}
\end{picture}
\caption{Check of the independence of the likelihood with respect to $\tan
  \beta$ and $A_0$ in the CMSSM fits. (a) $A_0$ and $\tan \beta$ values of
  the sampled points, (b) binned distribution of $D
  \chi^2$, the difference between the $\Delta \chi^2$ calculated by our
  $A_0-\tan   \beta$ independent approximation and the $\Delta \chi^2$
  calculated by event generation. \label{fig:check}} 
\end{center}\end{figure}
We now wish to check our approximation that the likelihood ${\mathcal L}$ is
independent of the other CMSSM parameters in the fit aside from $m_0$ and
$M_{1/2}$. In order to do this, we first take one hundred different sampled
points at random from the 2.7 million different points in the fits, before the
likelihood has been re-weighted by the $\alpha_T-$search. Each point
has different Standard Model and CMSSM parameters, and the probability of
selecting it is proportional to its posterior probability density of the fit
to indirect data. These points therefore represent a faithful sampling from
the fit, 
and we may estimate different quantities by calculating their probability
distributions from this faithful sampling. We show, for reference, the values
of $A_0$ and $\tan \beta$ of the sampled points in Fig.~\ref{fig:check}a.
For each of these points, we generate 10 000 SUSY signal events and estimate
the numbers of events passing cuts in the $\alpha_T$ search, as in
section~\ref{sec:CMS}, turning the number of events into a $\Delta \chi^2$ for
the search.  
We then compare this $\Delta \chi^2$ obtained by simulating events with the one
from the linear interpolation of $A_0=0$, $\tan \beta=3$ in
Fig.~\ref{fig:scan}a. The difference between the two, $D \chi^2$, measures how
much our approximation is violated. 

In Fig.~\ref{fig:check}b, we bin in 
$D \chi^2$, showing how many of the 100 points lie in each bin. Up to an
overall normalization of 1/100 then, this is an estimate of the probability
distribution of $D\chi^2$ in our fit.
All of the 100 points sampled have $D\chi^2<0.2$ except for six, which are shown by
asterisks in Fig.~\ref{fig:check}a. The most egregious 
has $D \chi^2=3.7$, but is heavily ruled out by the $\alpha_T$-search, with a
$\Delta \chi^2$ of 15. The others all have $m_0<100$ GeV,
$m_{1/2} \sim 300$ GeV, $\tan \beta \in (5,10)$ and $A_0 \in (-0.5, -1.0)$
TeV. The fact that the egregious points are clustered in parameter space can
be understood as follows.
The large values of $-A_0$ at small $m_0$ and $m_{1/2}$ significantly change
the up and down squark masses at these 
points compared to $A_0=0$ via the MSSM renormalisation group
equations (RGEs)\footnote{The trilinear couplings enter the RGEs for the SUSY
 breaking masses of the Higgs bosons, which then enter the RGEs of the first
 two families of squarks.}, having a  
resultant effect on the overall cross-section and hence on the number of SUSY
signal events passing cuts. 
Negative $A_0$ is preferred by the global fit as compared to positive $A_0$
because it tends to give larger lightest CP-even Higgs masses by enhancing
the stop mixing and enhancing the (dominant) stop loop correction to the Higgs
mass. Once small $m_{1/2}$ has been selected, the global fits
preferentially select $\tan \beta \in (5,10)$, as Fig.~\ref{fig:m12tb}a
shows. 
The anomalous magnetic moment of the muon prefers
these values of $\tan \beta$ for such light sparticles. 
For small $m_0$ and $m_{1/2}$, $A_0 < -1$ TeV results in negative stau mass
squared parameters, resulting in a charge breaking minimum of the scalar potential. Such values are
therefore disallowed in the global fit. 

We conclude that the assumed independence to parameters other than
$m_0$ and $m_{1/2}$ is quite a good approximation, since, excluding the
most egregious point (which is ruled out anyway), we obtain
$D\chi^2=0.04\pm0.22$.  

\acknowledgments
This work has been partially supported by STFC\@. We thank other members of the
Cambridge SUSY working group for discussions held, particularly C Lester and S
Williams on $\alpha_T$. We thank D Grellscheid and P Richardson for
communication regarding {\tt Herwig++}, and F Moortgat and K Matchev for
providing details of the CMS analysis.
\bibliographystyle{JHEP}
\bibliography{a}

\end{document}